\documentclass[runningheads]{cl2emult}

\usepackage{makeidx}  
\usepackage{graphicx} 
\usepackage{subeqnar} 
\usepackage{multicol} 
\usepackage{cropmark} 
\usepackage{eso}      
\makeindex            

\def\capt{\small \baselineskip 12pt }
\def\be{\begin{equation}}
\def\ee{\end{equation}}
\def\se#1{\S\ref{sec:#1}}
\def\fig#1{Figure~\ref{fig:#1}}
\def\equ#1{Eq.~(\ref{eq:#1})}

\def\bc{{\bf c}}
\def\bd{{\bf d}}
\def\br{{\bf r}}

\def\etal{{\it et al. }}

\def\eg{{\it e.g.}}

\def\apriori{{\it a priori }}

\def\kms{\,{\rm km\,s^{-1}}}
\def\hmpc{\,{\rm h^{-1}Mpc}}
\def\hkpc{\,{\rm h^{-1}kpc}}
\def\ihmpc{\,(h^{-1} {\rm Mpc})^{-1}}

\def\3hmpc{\,(h^{-1} {\rm Mpc})^3}
\def\ln{{\rm ln}}
\def\log{{\rm log}}

\def\la{\langle} \def\ra{\rangle}

\def\omm{\Omega_m}
\def\oml{\Omega_\Lambda}
\def\omb{\Omega_b}

\def\lcdm{$\Lambda$CDM}

\def\h60{h_{60}}

\def\L{{\cal L}}

\def\kb{k_{\rm b}}
\def\sigv{\sigma_{\rm v}}
\def\rc{r_{\rm c}}

\begin{document}
\title*{Nonlinear Peculiar-Velocity Analysis and PCA}
\toctitle{Nonlinear Peculiar-Velocity Analysis and PCA}
%
\titlerunning{Nonlinear Velocity Analysis and PCA}
%
\author{Avishai Dekel\inst{1}
\and Amiram Eldar\inst{1}
\and Lior Silberman\inst{1}
\and Idit Zehavi\inst{2}}
\authorrunning{Avishai Dekel et al.}
%
\institute{Racah Institute of Physics, The Hebrew University, Jerusalem 91904,
Israel
\and 
NASA/Fermilab Astrophysics Group, FNAL, Box 500, Batavia, IL 60510, USA}

\maketitle              

\begin{abstract}
We allow for nonlinear effects in the likelihood analysis of peculiar
velocities, and obtain $\sim 35\%$-lower values for the cosmological 
density parameter and for the amplitude of mass-density fluctuations.
The power spectrum in the linear regime is assumed to be of the flat 
\lcdm\ model ($h=0.65$, $n=1$) with only $\omm$ free.  
Since the likelihood is driven by 
the nonlinear regime, we ``break" the power spectrum at $\kb\sim 0.2\ihmpc$
and fit a two-parameter power-law at $k>\kb$. This allows for an unbiased 
fit in the linear regime.  Tests using improved mock catalogs demonstrate 
a reduced bias and a better fit.  We find for the Mark III and SFI data
$\omm=0.35 \pm 0.09$ with $\sigma_8 \omm^{0.6}=0.55 \pm 0.10$ 
(90\% errors).
When allowing deviations from \lcdm, we find an indication for a wiggle 
in the power spectrum in the form of an excess near $k\sim 0.05$ and a 
deficiency at $k \sim 0.1\ihmpc$ --- a ``cold flow" which may be related 
to a feature indicated from redshift surveys and the second peak in the 
CMB anisotropy.
A $\chi^2$ test applied to principal modes demonstrates that the 
nonlinear procedure improves the goodness of fit.  The Principal Component 
Analysis (PCA) helps identifying spatial features of the data and 
fine-tuning the theoretical and error models.  We address the 
potential for optimal data compression using PCA.
\end{abstract}

\section{Introduction}
\label{sec:intro}

When the large-scale density fluctuations are assumed to be Gaussian,
they are fully characterized by their power spectrum $P(k)$,
which is directly related to cosmological parameters that we wish
to estimate.  The $P(k)$ as estimated from redshift surveys 
is contaminated by unknown ``galaxy biasing", redshift distortions,
triple-value zones and the nonlinearity of the density field.
It is therefore advantageous to estimate the mass-density $P(k)$ directly from 
dynamical data such as peculiar velocities (PVs), which avoid biasing and
suffers from weaker nonlinear effects.
The $P(k)$ is evaluated here via a likelihood analysis of the individual PVs, 
not via a reconstruction method like POTENT (Dekel \etal 1999).  
The simplifying assumptions made in turn are that the PVs are drawn from a 
Gaussian random field, the velocity correlations can be derived
from the density $P(k)$ using linear theory, and the 
errors are Gaussian. The method assumes as prior a parametric
functional form for the $P(k)$, which allows cosmological parameter 
estimation.

We use two PV catalogs.
The Mark III (M3) catalog (Willick \etal 1997) contains $\sim\!3000$ 
galaxies within $\sim\!70 \hmpc$ from several datasets of spirals 
and ellipticals with distances inferred by the Tully-Fisher (TF)
and $D_n\!-\!\sigma$ methods.  
The SFI catalog (Haynes \etal 1999) consists of $\sim\!1300$ late-type 
spirals with I-band TF distances from two datasets.  The catalogs cover 
a similar volume, with the M3 sampling denser nearby and sparser at large 
distances.  The errors are $\sim 15-21\%$ of the distance.
In M3, the galaxies were grouped into $\sim 1200$ objects and then
corrected for Malmquist bias.  The SFI data were corrected for  
biases as in Freudling \etal (1999, F99).

In Zaroubi \etal (1997, Z97) and F99, we applied a likelihood analysis 
with a {\it linear\,} $P(k)$ model on all scales, taken from 
the family of COBE-normalized CDM models.  The free parameters were 
$\omm$, $n$, and $h$.
The two datasets yielded a high $P(k)$, defining a 
surface in parameter space:  
$\omm\, h_{65}^{1.3}\, n^2 \simeq 0.54 \pm 0.10$ 
(flat universe, no tensor fluctuations).  Correspondingly, 
$\sigma_8 \omm^{0.6} \simeq 0.84 \pm 0.12$.
These seemed to be consistent with the $2\sigma$ lower bounds of 
$\omm > 0.3$ obtained from PVs by other biasing-free methods
(Nusser \& Dekel 1993; Dekel \& Rees 1994; Bernardeau \etal 1995),
but they imply higher values than obtained from other estimators, e.g. 
based on cluster abundance (White, Efstathiou \& Frenk 1993; Eke \etal 1998).

The method has been tested using mock catalogs drawn from an 
N-body simulation of limited resolution.  We therefore generated 
new mock catalogs based on simulations of higher resolution, which
reveal a significant bias in the linear analysis.
The fit is driven by the small scales, because close pairs tend to consist 
of nearby galaxies with small errors, and therefore weak nonlinear effects
on small scales may bias the results.
We improve the analysis by adding free parameters that allow 
independent matching of the nonlinear behavior and thus
an unbiased determination of the linear part of the spectrum
and the associated cosmological parameters.
We then investigate the $P(k)$ independent of a specific cosmological model,
by allowing as free parameters the actual values of $P(k)$
in finite intervals of $k$ (also in Zehavi \& Knox, in preparation).  

The likelihood analysis provides only relative likelihoods of the parameters, 
not an absolute goodness of fit (GOF).  An indication for a problem 
in the GOF of the linear analysis came from a $\chi^2$ estimate in modes
(Hoffman \& Zaroubi 2000). It seems to be associated with a problem noticed 
earlier by F99, of a spatial gradient in the obtained value of $\omm$.
We therefore develop a method based on $\chi^2$ in PCA as a tool for 
evaluating the GOF in our new procedure compared to the old one.

In \se{method} we describe the method. 
In \se{mock} we test and calibrate it using mock catalogs.
In \se{broken} we present the resultant $P(k)$ and
$\omm$ for \lcdm. 
In \se{bins} we detect hints for deviations from this model.
In \se{pca} we describe the PCA.
In \se{gof} we evaluate GOF via $\chi^2$ in modes of PCA.
In \se{conc} we conclude.

\section{Method}
\label{sec:method}

As explained in Z97 and F99, the goal is to estimate the density $P(k)$ 
from PVs by finding maximum likelihood values for 
parameters of an assumed model $P(k)$.
Under the assumption that the underlying velocities and the
observational errors are independent Gaussian random fields,
the likelihood is 
\be
\label{eq:like}
\L = {1 \over [ (2\pi)^n \det(C)]^{1/2} }
  \exp\left( -{1\over 2}\sum_{i,j}^n {u_i C_{ij}^{-1} u_j}\right)\,,
\ee
where $\{u_i\}_{i=1}^{n}$ are the observed PVs
at locations $\{\br_i\}$, and $C$ is their correlation
matrix. Expressing each data point as signal plus noise, 
$u_i=s_i+n_i$, 
\be
C_{ij}\equiv \la u_i u_j \ra = \la s_i s_j \ra + \la n_i n_j \ra
\equiv S_{ij}+ N_{ij} \ .
\label{eq:C}
\ee
$S$ is the correlation matrix of the signal,
calculated from the $P(k)$ model at $\br_i$, and $N$ is the error 
matrix, assumed diagonal because the errors of the objects are uncorrelated.
For a given $P(k)$, $S_{ij}$ are calculated via the parallel and perpendicular
velocity correlation functions, $\Psi_{\Vert}$ and $\Psi_{\perp}$,
\be
S_{ij}=\Psi_{\perp}(r)\sin\theta_i \sin\theta_j +
\Psi_{\Vert}(r)\cos\theta_i
\cos\theta_j  \, ,
\ee
where $r=\vert \br \vert=\vert \br_j-\br_i \vert$ and the angles are
defined by $\theta_i=\hat{\br_i}\cdot\hat{\br}$ (G\'orski 1988; Groth,
Juszkiewicz \& Ostriker 1989).  In linear theory, 
\be
\Psi_{\perp,\Vert}(r)= {H_0^2 f^2(\omm )\over 2 \pi^2}
\int_0^\infty P(k)\, K_{\perp,\Vert}(kr)\, dk \,,
\ee
where $K_{\perp}(x) = j_1(x)/ x$ and $K_{\Vert}(x) = j_0-2{j_1(x)/
x}$, with $j_l(x)$ the spherical Bessel function of order $l$.  The
$\omm$ dependence enters as usual via
$f(\omm)\simeq \omm^{0.6}$.

For each choice of parameters, $C$ is computed, inverted, and substituted in 
\equ{like}. Exploring parameter space, we find the parameters that maximize 
the likelihood.  The main computational effort is the inversion of 
$C$ in each evaluation of the likelihood. It is an $n \times n$ matrix, 
where currently $n\sim 10^3$, but when $n$ increases soon,  
a procedure for data compression will be required (see \se{pca}).

The measurement errors add in quadrature to the $P(k)$ and thus
propagate into a systematic uncertainty.  Z97 used \apriori\ estimates 
of the errors, while F99 incorporated the errors into the likelihood
analysis via an error model with free parameters.
They found errors within 5\% of the \apriori\ estimates, thus allowing 
the use of the \apriori\ estimates here.
Relative confidence levels are estimated by approximating $-2\ln\L$ as a
$\chi^2$ distribution with respect to the model parameters.  

In the linear regime, we use as prior the parametric form 
based on the general CDM model,
$ P(k) = A_{\rm c}(\omm,\oml,n)\ T^2(\omm,\omb,h; k)\ k^n$,
where $A_{\rm c}$ is the normalization factor and
$T(k)$ is the CDM transfer function proposed by Sugiyama (1995).
We restrict ourselves in the present paper to 
flat cosmology with a cosmological constant ($\omm+\oml=1$), 
a scale-invariant power spectrum on large scales ($n=1$),
and a Hubble constant $h=0.65$.
The baryonic density is set to be $\omb=0.024 h^{-2}$ (Tytler, Fan \&
Burles 1996), and the amplitude $A_{\rm c}$ is fixed by the COBE 4-year
data (Hinshaw \etal 1996; G\'orski \etal 1998).
The analysis has been repeated with a later estimate of
$\omb=0.019 h^{-2}$ (Burles \etal 1999) and an alternative COBE
normalization (Bunn \& White 1997), with variations $<2\%$
due to these changes.

An accurate nonlinear correction to the linear
{\it velocity\,} power spectrum could have been very useful
in avoiding the bias associated with the linear analysis, but such a 
correction is not yet available.
An empirical approximation does exist for the nonlinear
correction to the {\it density\,} $P(k)$ (Peacock \& Dodds 1996, PD),
but the generalization to a velocity correction is not straightforward.
Here, we detach the nonlinear regime
from the linear regime by a ``break" in the $P(k)$ at
a wavenumber $\kb$. We then assume the \lcdm\ shape at $k<\kb$,
determined by $\omm$, and allow an almost
arbitrary power law with two free parameters to fit the data at $k>\kb$.
This ``feeds the nonlinear monster" while the linear part of the
spectrum, and the associated cosmological parameters, are freed
to be determined unbiased.

\section{Testing with Mock Catalogs}
\label{sec:mock}

We test the method using mock catalogs based on
cosmological simulations in which the ``true" cosmological parameters
and linear $P(k)$ are fully known \apriori, and where nonlinear effects
are simulated with adequate accuracy.
We use the unconstrained ``GIF" simulation (Kauffmann \etal 1999a)
of the flat \lcdm\ cosmology with $\omm=0.3$.
The initial fluctuations were Gaussian, adiabatic and scale-invariant,
$n=1$. The $P(k)$ shape parameter was $\Gamma=\omm h=0.21$
(namely $h=0.7$) and the amplitude is such that $\sigma_8=0.9$.
consistent with both the present
cluster abundance and COBE's measurements on large scales.
The $N$-body code is a version of the adaptive particle-particle
particle-mesh (AP$^3$M) Hydra code developed as part of the VIRGO
supercomputing project (Jenkins \etal 1998).
The simulation has $256^3$
particles and $512^3$ cells, and a gravitational softening
length of $30 \hkpc$, inside a box of side $141.3\hmpc$.
Dark-matter halos were identified using
a friends-of-friends algorithm with a linking length of 0.2
and a minimum of 10 particles per halo was imposed.
Luminous galaxies were planted in the halos based on a semi-analytic
scheme (Kauffmann \etal 1999a, 1999b).
We assigned to each galaxy a linewidth based on the TF relation and
the scatter assumed in Kolatt \etal (1996).
We then generated 10 mock catalogs which resemble the M3
catalog, with random distance errors and random sampling.
The mock data were grouped and corrected for Malmquist biases
exactly as in the real data. 

\begin{figure}[t!]
\vspace{5.8truecm}
\includegraphics{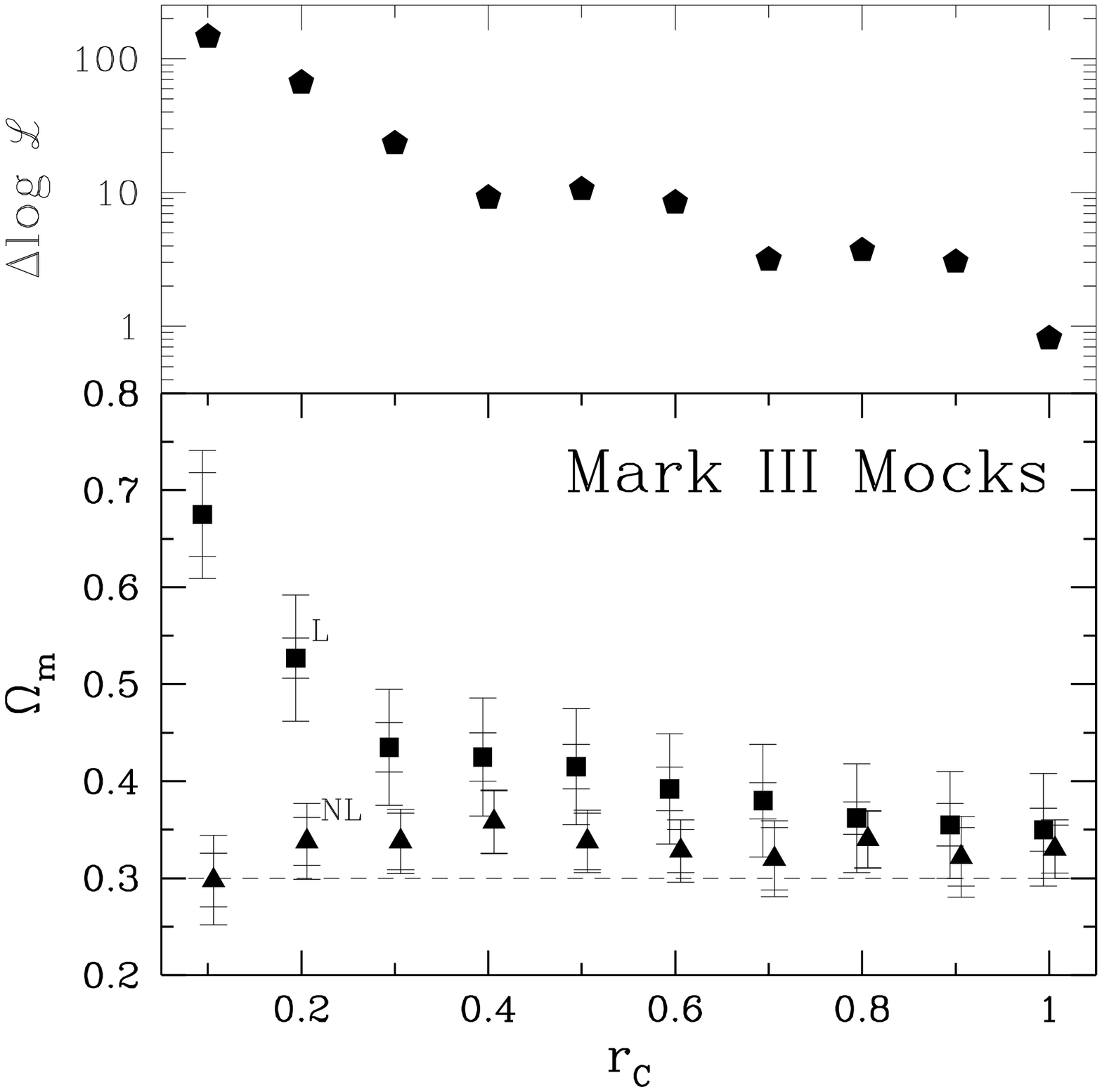}
 \includegraphics{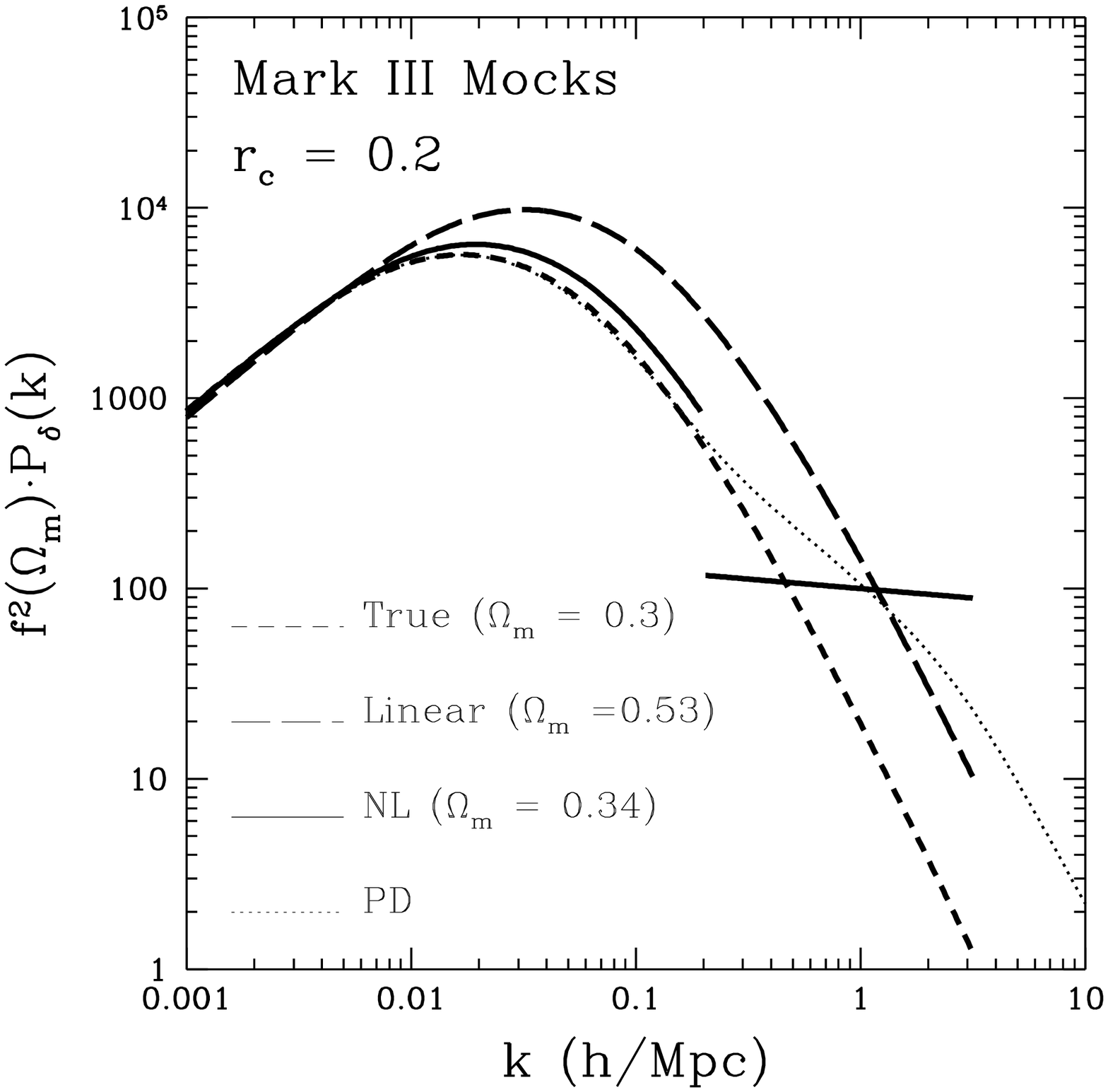}
\caption{\protect \capt
Testing the method.
{\bf Left:} 
{\it Bottom:} the recovered $\omm$ as a function of the degree of linearity 
of the dataset, $\rc$.
Each symbol marks the average over 10 mock M3 catalogs,
and the inner and outer error bars mark the scatter
and $90\%$ likelihood uncertainty. 
The squares are the results of the linear analysis;
they show a bias that
increases with the degree of nonlinearity.
The triangles are the results of the 
broken-\lcdm\ $P(k)$; the bias is drastically reduced.
{\it Top:} the corresponding improvement in $\log \L$
for the nonlinear versus linear analysis.  
{\bf Right:}
Mean power spectra recovered from the M3 mock catalogs of $\rc =0.2$.
The target is the true linear $P(k)$.  The biased linear result is shown.
The result from the nonlinear analysis with $\kb=0.2\ihmpc$ is marked ``NL".  
The $P(k)$ is in units of $\!\3hmpc$.
}
\label{fig:mock}
\end{figure}

To study how the method performs in the presence of nonlinear effects,
we produced a suite of 10 sets of 10 mock catalogs each,
spanning a range of degree
of nonlinearity, created by varying the criterion for the exclusion of
cluster galaxies. Galaxies were excluded if they lie within a distance
$\rc$ from the cluster center. The ``linearity parameter" $\rc$ is
measured in units of $3.5\hmpc$ and $1.5\hmpc$ for spirals and ellipticals
respectively, and it ranges from $\rc=0.1$ to $1$ in steps of $0.1$.
The likelihood analysis has been applied to each of the $10 \times 10$ mock
M3 catalogs. The recovered values of $\omm$ are shown in \fig{mock} (left).
The ``true" target value is the $\omm =0.3$ of the simulation.

We first apply the purely linear analysis, with the linear \lcdm\ power
spectrum at all scales,
and with $\omm$ as the only free parameter.
We see that the linear likelihood analysis systematically overestimates the
value of $\omm$. As the data become more nonlinear, the recovered
value of $\omm$ becomes higher, and the bias more significant.
Next, we apply the improved procedure, allowing for a break in the $P(k)$
at $\kb=0.2\ihmpc$ and additional free parameters in the nonlinear
regime. We see that
the bias is practically removed for all levels of nonlinearity.
The figure also shows the corresponding improvement in $\log \L$
when the linear analysis is replaced with the nonlinear analysis.
\fig{mock} shows the corresponding power spectra
from the M3 mock catalogs of linearity $\rc =0.2$.

Our conclusion from the above test using the mock catalogs is that,
in the presence of significant nonlinear effects in the data,
the purely linear likelihood analysis might yield
a biased estimate of $P(k)$ and $\omm$.
The broken-$P(k)$ analysis successfully eliminates the dependence
of the results on the nonlinear effects and practically
corrects the bias in the results.

\section{Broken \lcdm: the Value of $\omm$}
\label{sec:broken}

\begin{figure}[t!]
\vspace{5.8truecm}
\includegraphics{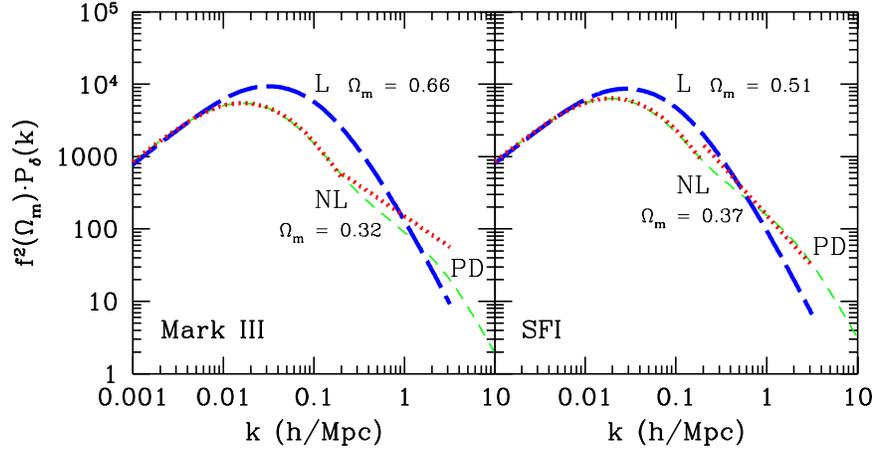}
\caption{\protect \capt
The recovered power spectra from the real data.
The $P(k)$ yielded by the linear analysis is marked ``L",
while the nonlinear analysis, with a break at $k=0.2\ihmpc$,
is marked ``NL".
}
\label{fig:m3sfi}
\end{figure}

\begin{figure}[t!]
\vspace{4.7truecm}
\includegraphics{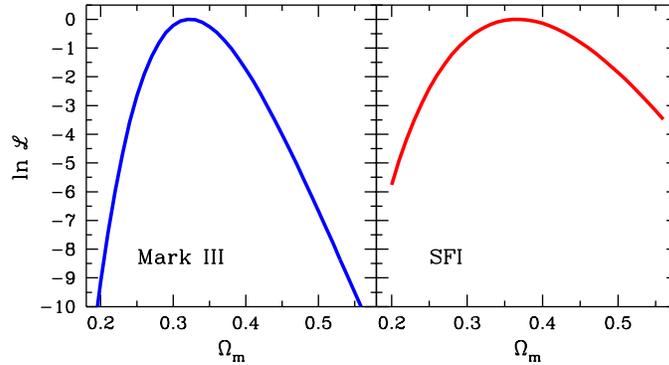}
\caption{\protect \capt
Likelihood function for the values of $\omm$ due to the nonlinear
analysis of the real data.
}
\label{fig:likeomega}
\end{figure}

We now analyze the real data.
Our \lcdm\ model is restricted again to a flat universe with $h=0.65$,
$\omb h^2 =0.02$ and $n=1$, leaving only one cosmological parameter
free, $\omm$.  Unlike the mock tests, we now do not know  
that \lcdm\ is the right model or that the values of the fixed parameters
are accurate.  
\fig{m3sfi} shows the resultant power spectra.
The linear analysis yields a high $P(k)$ amplitude and a high value of
$k_{\rm peak}$, corresponding to $\omm=0.56\pm 0.04 $ and
$\omm=0.51\pm 0.05$ for M3 and SFI ($90\%$ errors), 
consistent with Z97 and F99 for the fixed values of $h$ and $n$
quoted above.
The nonlinear analysis yields a shift of $k_{\rm peak}$ towards lower $k$'s,
associated with lower values of 
$\omm=0.32\pm 0.06$ and $\omm=0.37\pm 0.09$ for M3 and SFI.
\fig{likeomega} shows the likelihood as a function of $\omm$, where for each
value of $\omm$, the nonlinear parameters obtain their most likely values.
When we marginalize over the nonlinear parameters, the likelihood function
is very similar.
The corresponding best-fit values of $\sigma_8\omm ^{0.6}$ are $0.49 \pm 0.06$
and $0.63 \pm 0.08$ for M3 and SFI.  These values are consistent with the 
estimates from cluster abundance (\eg, Eke \etal 1998).

The reduction in the value of $\omm$ due to the nonlinear correction
is similar to the mock catalog case
at a relatively high degree of nonlinearity, $\rc \simeq 0.2$ in
\fig{mock}.
The power-law segments roughly coincide
with the linear \lcdm\ segments at $\kb$, indicating that this broken
$P(k)$ is a sensible approximation to the actual shape of $P(k)$.
The two catalogs basically yield consistent results.  As expected, the
nonlinear correction for M3 is larger than for SFI, because the former
has more galaxies nearby in close pairs with small errors.
The likelihood improvement for M3 is very significant, 
$\Delta \ln \L \simeq 22$, while for SFI it is moderate, 
$\Delta \ln \L \sim 2.8$.

We find that the results are quite insensitive to the choice of $\kb$
over a wide range.  At $\kb < 0.1$, corresponding to large pair separation
and thus involving mostly distant objects of large errors, there are
insufficient data to constrain the power spectrum, and the errors become big.
At very large values of $\kb$, the analysis recovers the
results of the linear analysis, but only when $\kb$ approaches the artificial
cutoff applied to $P(k)$ arbitrarily at $k_{\rm max}=8\ihmpc$
for the purpose of finite numerical integration.
It seems that any little freedom allowed in the model beyond the strict
linear $P(k)$ is enough for correcting the bias associated with the
linear analysis.
Indeed, an alternative way to incorporate nonlinear effects is by adding to the
linear velocity correlation model a free parameter of uncorrelated velocity
dispersion at zero lag, $\sigv$.  When this correction is applied by itself,
the best value of $\omm$ becomes $0.38$ (instead of 0.56 in the linear
analysis) with $\sigv=250\kms$.  When the two different nonlinear corrections
are applied together, a break plus $\sigv$, the best-fit value of $\sigv$
is close to zero, indicating that the two corrections are practically
redundant.

\section{Deviations from \lcdm}
\label{sec:bins}

We now push the idea further, and divide the power spectrum into 4 detached 
segments.  This allows a more general shape for $P(k)$, less dependent on 
\apriori\ assumptions about a physical model such as \lcdm, but at the 
expense of giving up the attempt to determine cosmological parameters.
(The choice of a series of independent step functions, or ``band powers"
forming a histogram, is especially appealing computationally, because it
makes the correlation matrix a linear combination of 
the correlation matrices of the individual segments, and then the 
integrals involved need to be computed only once.)
Our 5-parameter model actually consists of the following segments: 
(a) COBE-normalized \lcdm\ in the extreme linear regime,
$k \leq 0.02$, with one free parameter, $\omm$,
(b) a free constant amplitude in the interval $0.02<k\leq0.07$ near 
$k_{\rm peak}$,
(c) an independent free amplitude in the interval $0.07<k\leq0.2$,
and (d) a power law as before in the nonlinear regime, $k>0.2$.

\begin{figure}[t!]
\vspace{5.6truecm}
\includegraphics{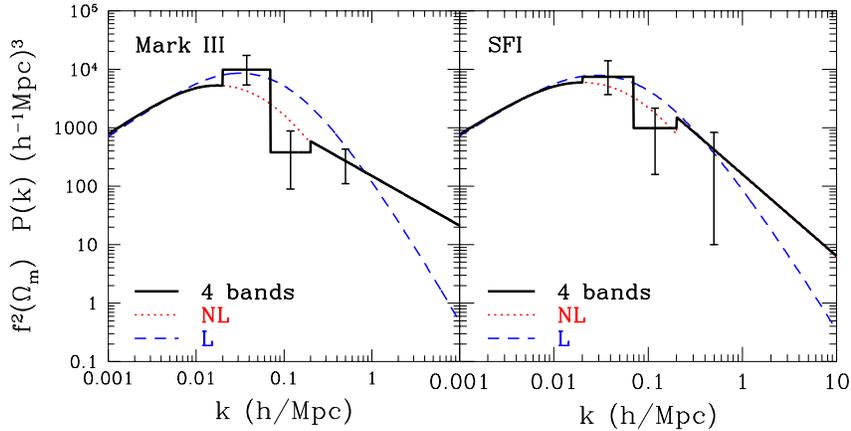}
\caption{\protect \capt
The 4-band power spectrum of the real data,
compared to the best-fit \lcdm\ power spectra, linear (L) and nonlinear
with a break (NL).
}
\label{fig:4steps}
\end{figure}

\fig{4steps} shows the recovered 4-band $P(k)$, in comparison with the 
\lcdm\ results.  The nonlinear segment practically recovers the
results of the broken-\lcdm\ analysis.  The two most linear segments 
lie along the results of the linear analysis, with a higher peak than
in the broken-\lcdm\ case.  The most interesting feature is
the low amplitude in the interval $(0.07,\,0.2)$,
in the ``blue" side of the peak and just shy of
the transition to the nonlinear regime.
The features in the linear regime contribute only a marginal improvement
to the overall likelihood, and should therefore be considered as a marginal 
hint only.  This could be a fluke due to distance errors and cosmic variance,
but the consistent appearance in the two datasets makes it intriguing.

The marginal deviation from the broken-\lcdm\ $P(k)$ thus consists
of a wiggle, with a power excess near the peak, $k\sim 0.05$,
and a deficiency at $k \sim 0.1$.
The missing power is reminiscent of other indications for ``cold flow"
in the peculiar velocity field in the local cosmological neighborhood.
While the streaming motions on $\sim 30\hmpc$ scales are $\sim 500\kms$,
the dispersion velocity of field galaxies is only $\sim 200\kms$, indicating 
a high Mach number (\eg, Suto, Cen \& Ostriker 1992; 
Chiu, Ostriker \& Strauss 1998; Dekel 2000).

\section{Principal Modes: S+N versus S/N}
\label{sec:pca}

The likelihood analysis is limited to estimating
relative likelihoods for the parameters,
but it does not address the absolute GOF of the model and data.
The linear analysis of both M3 and SFI revealed a worry concerning the GOF, 
in the sense that when applied separately to two halves of the
data, separated either by distance or line width,
the distant data prefer a lower $\omm$.
The mock catalogs have not revealed a similar problem, indicating that it is
caused by inadequacies of the correlation matrix, associated with either 
the theoretical or the error model.
This motivates an attempt to evaluate GOF,
hoping that the broken-$P(k)$ resolves the two-halves problem.

Assume a data vector $\bc$, which is a random realization of an
Gaussian distribution, with the correlation
matrix $C=\langle \bc \bc^{\dagger }\rangle$.
A global GOF could be evaluated using the $\chi^2$ statistic,
$\chi^{2} = \bc ^{\dagger} C^{-1} \bc$.
If $C$ is the true correlation matrix, then this value should
obey a $\chi^2$ distribution with $n$ degrees of freedom. But this
single number cannot capture all the particulars of the fitting process.
A PCA, in which the data are represented
in terms of the eigenvector basis of the (assumed) correlation matrix,
provides a powerfull tool
for identifying gross features of the data and model,
and for evaluating GOF in fine detail.
We diagonalize the matrix via the transformation $\bd=\Psi \bc $, 
into $D=\langle \bd \bd^\dagger \rangle =\Psi C\Psi ^{\dagger}$.
The likelihood analysis can be performed in terms of the new ``data"
points $\bd$.
The rows $d_i$ of $\Psi$ are the eigenvectors, or the principal 
modes, and the diagonal terms $\lambda_i$ of $D$ are the corresponding
eigenvalues.  The new variables, $d_i/\sqrt{\lambda _i}$, 
are expected to be uncorrelated unit Gaussian random variables.  This property
is a measure of GOF, which can be evaluated by the $\chi^{2}$ statistic.
If this test uncovers systematic effects, they may be
associated with certain features of the data and model
via a detailed investigation of the eigenmodes.

\begin{figure}[t!]
\vspace{7.8truecm}
\includegraphics{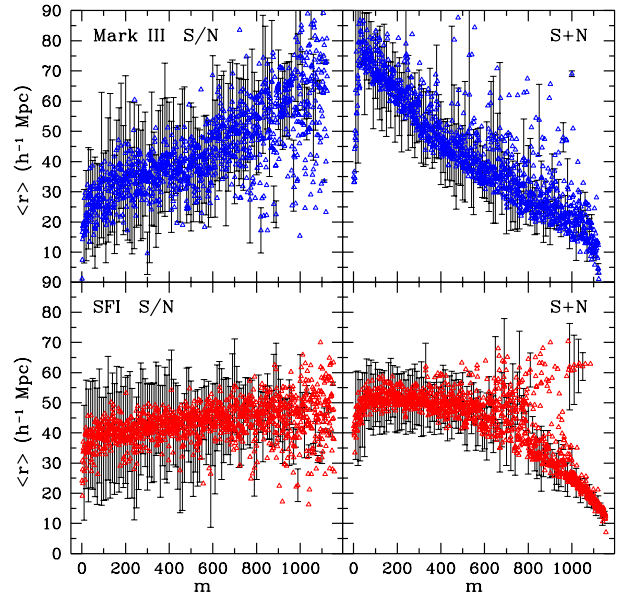}
\caption{\protect \capt
Average distances associated with the eigenmodes of the 
linear \lcdm\ model.
The eigenmodes are ranked by decreasing eigenvalues
(low $m$ --- high eigenvalue).
The standard deviation is shown for every 10th mode.
}
\label{fig:r}
\end{figure}

The eigenmodes are ordered by the amplitude of their eigenvalues,
from large to small. The high-eigenvalue modes are assigned a higher
signifcance, because the confidence levels in the recovered parameters 
inversly correlate with the squares of the eigenvalues (Tegmark, Taylor 
\& Heavens 1997), and because perturbation analysis implies that 
small-eigenvalue modes are more sensitive to perturbations in the 
correlation matrix.  Since our correlation matrix is expected to
be only an approximation, it would be advantageous to avoid these modes. 
A straightforward application of PCA is with the standard correlation
matrix, $C=S+N$, where large eigenvalues may correspond
to large signal or large noise or both.
Another possibility, which we term $S/N$,
is to first perform a ``whitening" transformation,
$\bd=N^{-\frac{1}{2}}\bc$ (Vogeley \& Szalay 1996),
such that the new correlation matrix is
$D=N^{-\frac{1}{2}}SN^{-\frac{1}{2}}+I$,

The eigenmodes can help us identify certain features of the data and models.
For example, the distance associated with a mode may be useful  
because distant modes are typically noisier and because it may connect
to the two-halves problem.  For eigenvector $v$, the average distance is
$\langle r\rangle _{v}=\sum_g |v(g)|^{2} r(g)$,
where the sum is over the sample of galaxies, $r(g)$ is the distance
of galaxy $g$, and $v(g)$ defines the vector $v$ in the basis $g$.
If the standard deviation, defined in analogy, is small compared to the 
average distance, then most of the information associated with this mode
comes from galaxies within a certain distance range.
\fig{r} shows the average distance for each mode.  The robust
high S/N modes are typically associated with nearby data, which are
of smaller errors.  These tend to involve close pairs,
and therefore stronger nonlinear effects, which makes the
nonlinear correction a must.
The S+N modes show a correlation with distance
in the opposite sense, in which the high-eigenvalue modes
are typically associated with large distances and therefore
noisy data. This means that most of the S+N modes are dominated by the
noise, which would not allow a sensible truncation by S+N modes,
but should allow a more sensitive measure of GOF, refering in
particular to the error model.

\section{Goodness of Fit Mode by Mode}
\label{sec:gof}

\begin{figure}[t!]
\vspace{7.5truecm}
\includegraphics{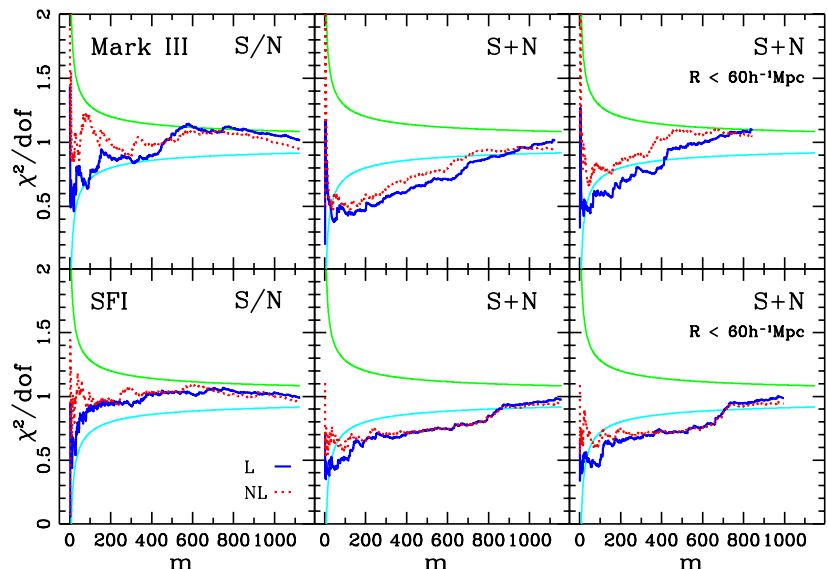}
\caption{\protect \capt
Cummulative $\chi ^2/dof$ as a function of mode number
for the linear \lcdm\ model (solid) and for the broken-\lcdm\ $P(k)$ (dotted).
The two smooth lines mark the expected $2\sigma$ deviations.
}
\label{fig:chi2_cum}
\end{figure}

After PCA, we expect $\chi ^2_i = d_i^2/\lambda _i$
to be about unity for each mode separately.
This tests whether the eigenmodes of the prior correlation matrix
are really uncorrelated with the proper variance,
and, in the case of a poor fit for a certain mode, it can guide us to
the source of the problem.
The statistic shown in \fig{chi2_cum}, for each mode number $m$, 
is the cummulative $\chi^2$ per degree of freedom, $\sum ^m_{i=1}\chi _i^2/m$. 
The expected value is unity, and the expected standard deviation is
$\sim\sqrt{2/m}$.

For the S/N PCA of M3, the GOF of the linear model is marginal,
in the sense that the typical deviations are at the $2\sigma$ level.
The low-$m$ modes, except for the very first ones, typically
have low $\chi^2/dof$ values, while the large-$m$ modes have high
values.
This behavior can be translated to a systematic
trend with distance via the correlation between distance and mode
(\fig{r}). It is therefore a reminiscence of the two-halves problem.
We see in \fig{chi2_cum} that
the broken-\lcdm\ model clearly improves the GOF as far as the
S/N modes of M3 are concerned, where the cummulative
$\chi^2/dof$
lies well inside the $2\sigma$ contours for all the modes, with no
apparent systematic dependence on $m$.  It implies that the
broken-\lcdm\ $P(k)$ is a more appropriate model for the data.

When we analyze the S+N modes in a similar way in \fig{chi2_cum},
the linear model, for both data sets,
shows a more severe deviation of $\chi^2/dof$ from unity, at the
$4-5\sigma$ level, and a similar systematic dependence on $m$.
The two-halves problem is very obvious here, with the more
distant, noisy data favoring a smaller amplitude for $P(k)$.
In this case, the use of the better, broken-\lcdm\ model makes only a
small improvement which does not resolve the problem.
This is a clear indication that something may be wrong in the error
model as well.

We recall that the low-$m$ S+N modes are associated with large
distances, where the errors are large and are known to a lesser accuracy.
Guided by \fig{r}, we try a poor-man compression of the data
by eliminating all the data points that lie at $>60\hmpc$.
This leaves us with $843$ out of the $1124$ (grouped) data points of M3,
and $996$ out of the $1156$ galaxies of SFI.
This truncation causes an increase in the best-fit value of $\omm$
by less than $3\%$, both for M3 and SFI,
and causes only a minor widening of the likelihood contours.
In the case of M3, we see in \fig{chi2_cum} that
the S+N modes of the linear model and the truncated data
show an improved GOF compared to the case of the whole data,
but the $\chi^2/dof$ still show $\sim 3\sigma$ deviations from unity
and a systematic dependence on $m$.
However, the S+N modes of the broken-\lcdm\ model and M3 data now do
lie within the $2\sigma$ contours.  The systematic trend with $m$ is
still apparent, indicating that the correlation matrix is still not
perfect;
either the error model is still only an approximation even for the
truncated data, or the broken-\lcdm\ $P(k)$ is not
yet a perfect model (as seen in \se{bins}), or
the signal and/or the noise are not exactly Gaussian, or all of the
above.

In the case of SFI,
while the S/N modes look very adequate with both models,
for the S+N cummulative statistic the improvements due to the nonlinear
correction and the elimination of large-distance galaxies are apparently
not enough for an acceptable GOF.
Since the large-eigenvalue S+N modes, which dominate the cummulative
statistic, are dominated by noise, the limited GOF is likely to point
at further shortcomings of the error model for SFI.

\section{Conclusion}
\label{sec:conc}

A likelihood analysis is supposed to recover unbiased values for the
free parameters of a model provided that the prior theoretical model
and the error model allow accurate description
of the data. We addressed here tools to recover the parameters given
incomplete knowledge of these models.

Using mock catalogs based on high-resolution simulations, we realized
that the likelihood analysis of PV data, based on the
linear \lcdm\ $P(k)$, is driven by the nonlinear part of the
spectrum which is not modeled accurately, and 
therefore yields biased results. 
A broken-\lcdm\ $P(k)$, in which the $k>\kb$
segment is replaced by a two-parameter power law,
allows a better, independent fit in the nonlinear regime. It then
frees the linear regime from nonlinear effects, and
yields unbiased results for $\omm$. The results are robust to the
specific choice of $\kb$; we choose $\kb = 0.2\ihmpc$, which is
where the nonlinear density $P(k)$ is expected to start deviating from
the linear $P(k)$ by the PD approximation.
The results are also robust to the exact way by which the nonlinear
effects are incorporated. When we add a zero-lag velocity dispersion
term to the correlation function, we obtain similar results.

When applied to the real data of M3 or SFI,
for a flat \lcdm\ model with $n=1$ and $h=0.65$,
the improved analysis yields best-fits of
$\omm = 0.32\pm 0.06$ and $0.37\pm 0.09$,
corresponding to
$\sigma _8\omm ^{0.6}\approx 0.49 \pm 0.06$ and $0.63 \pm 0.08$.
These are in agreement with most constraints
from other data, including CMB anisotropies and cluster abundance
(\eg, Bahcall \etal 1999). Joint analysis of PVs with other data is pursued
based on the linear analysis (Zehavi \& Dekel 1999) and the nonlinear
analysis (Bridle \etal 2000).

By allowing an more general shape for $P(k)$, with 4
detached segments, we detect 
a marginal deviation from the \lcdm, in the form of a
wiggle, with an enhanced amplitude near $k_{\rm peak} \sim 0.05$
and a significant depletion near $k \sim 0.1\ihmpc$.
This ``cold flow" on a scale of a few tens of megaparsecs
is reminiscent of similar indications from the $P(k)$
of galaxies and clusters in redshift surveys
(Baugh \& Gazta\~naga 1998; Landy \etal 1996; 
Einasto \etal 1997; Suhhonenko \& Gramann 1999).
Most recent is the wiggle indicated in the 2dF redshift survey.
The local phenomenon of cold flow may be related to the low second peak
as measured by Boomerang and Maxima in the CMB angular power spectrum
on a similar scale (Boomerang, de Bernardis \etal 2000; Maxima, Hanany \etal
2000).
The wiggly feature may be interpreted
as a deviation from the standard cosmological
mass mixture, \eg, a higher baryonic content than indicated by
the Deuterium abundance and Big-Bang nucleosynthesis, or a non-negligible
contribution from hot dark-matter.  But the excess required to
produce a significant wiggle seems to violate upper limits from
other data; the density of baryons is limited by He+D abundances
via the theory of Big-Bang nucleosynthesis (Tytler \etal 2000),
and the density of neutrinos is constrained by large-scale structure
(\eg, Ma 1999; Gawiser 2000).
The possibility that this feature is a statistical fluke due to cosmic
variance in the context of the \lcdm\ model cannot be ruled out yet.

A PCA, either in S/N or S+N modes, allows a
fine test of GOF, by applying a $\chi^2$ test mode by mode.
It shows that the broken-\lcdm\ model is a better fit to the data than
the purely linear \lcdm\ model.  For M3, using the ``whitened" S/N
modes, the nonlinear correction is enough to eliminate the ``two-halves"
problem that troubled the linear analysis.  When the S+N modes are analyzed,
the correction to the theoretical model is helpful but
not enough for an acceptable GOF.
When the M3 data is further truncated at $60\hmpc$, eliminating
distant galaxies for which the errors are large and the error model is
inaccurate, the GOF becomes acceptable.
For SFI, the S/N modes seem adequate, but the S+N PCA
indicates that the error model is still more complex than assumed.

The PCA is a powerful tool for addressing interesting properties
of the data and its relation to the theoretical and error model.
In particular, we associated each mode with 
a distance and learned about the correlation between mode eigenvalues
and distance errors. This was useful in the study of GOF and in
truncating the data to deal with inaccuracies in the error model.
The PCA will be extremely useful when one tries to compress the data
while keeping the optimal part for determining a specific desired
parameter.  This compression may be mandatory for computational reasons
when the body of data is excessively large.
Since the model is expected to be inaccurate or incomplete,
a proper compression of the data may in fact improve the results.
Such data compression using PCA in the context
of cosmic flows is in progress.

\section*{Acknowledgments}
This research has been partly supported by the Israel Science Foundation
grant 546/98,
by the US-Israel Binational Science Foundation grant 98-00217,
and by the DOE and the NASA grant NAG 5-7092 at Fermilab.
We acknowledge very stimulating discussions with Yehuda Hoffman, 
Lloyd Knox, Saleem Zaroubi and Amos Yahil.


 



\clearpage
\addcontentsline{toc}{section}{Index}
\flushbottom
\printindex

\end{document}